\documentclass[%
 aip,
cp,  
 amsmath,amssymb,
 reprint,%
]{revtex4-2}

\usepackage{graphicx}
\usepackage{dcolumn}
\usepackage{bm}

\usepackage[T1]{fontenc}

\usepackage{hyperref}
\usepackage{natbib}

\begin{document}

\title{Running Hubble constant \\ from the SNe Ia Pantheon sample?}

\author{Tiziano Schiavone}
    \affiliation{Department of Physics ``E. Fermi'', University of Pisa, Polo Fibonacci, Largo B. Pontecorvo 3, I-56127 Pisa, Italy\\
    tiziano.schiavone@phd.unipi.it}
    \affiliation{INFN, Istituto Nazionale di Fisica Nucleare, Sezione di Pisa, Polo Fibonacci, Largo B. Pontecorvo 3, I-56127, Pisa, Italy}
    
\author{Giovanni Montani}
    \affiliation{ENEA, Fusion and Nuclear Safety Department, C.R. Frascati, Via E. Fermi 45, I-00044 Frascati (Roma), Italy}
    \affiliation{Physics Department, ``Sapienza'' University of Rome, P.le Aldo Moro 5, I-00185 Roma, Italy}

\author{Maria Giovanna Dainotti}
    \affiliation{National Astronomical Observatory of Japan, 2 Chome-21-1 Osawa, Mitaka, Tokyo 181-8588, Japan\\
    maria.dainotti@nao.ac.jp}
    \affiliation{The Graduate University for Advanced Studies, SOKENDAI, Shonankokusaimura, Hayama, Miura District, Kanagawa 240-0193, Japan}
    \affiliation{Space Science Institute, Boulder, CO, USA}

\author{Biagio De Simone}
    \affiliation{Department of Physics ``E.R. Caianiello'', University of Salerno, Via Giovanni Paolo II, 132, Fisciano, I-84084 Salerno, Italy}
    \affiliation{INFN Gruppo Collegato di Salerno, Sezione di Napoli, c/o Dipartimento di Fisica “E.R. Caianiello”,
    Ed. F, Università di Salerno, Via Giovanni Paolo II, 132, Fisciano, I-84084 Salerno, Italy}

\author{Enrico Rinaldi}
    \affiliation{Physics Department, University of Michigan, Ann Arbor, MI 48109, USA}
    \affiliation{Theoretical Quantum Physics Laboratory, Center for Pioneering Research, RIKEN, 2-1 Hirosawa, Wako, Saitama 351-0198, Japan}
    \affiliation{Interdisciplinary Theoretical \& Mathematical Science Program, RIKEN (iTHEMS), 2-1 Hirosawa, Wako, Saitama, 351-0198, Japan}

\author{Gaetano Lambiase}
    \affiliation{Department of Physics ``E.R. Caianiello'', University of Salerno, Via Giovanni Paolo II, I-132-84084 - Fisciano, Salerno - Italy}
    \affiliation{INFN Gruppo Collegato di Salerno, Sezione di Napoli, c/o Dipartimento di Fisica “E.R. Caianiello”,
    Ed. F, Università di Salerno, Via Giovanni Paolo II, 132, Fisciano, I-84084 Salerno, Italy}

\date{\today} 

\begin{abstract}
The mismatch between different independent measurements of the expansion rate of the Universe is known as the Hubble constant ($H_0$) tension, and it is a serious and pressing problem in cosmology. We investigate this tension considering the dataset from the Pantheon sample, a collection of 1048 Type Ia Supernovae (SNe Ia) with a redshift range $0<z<2.26$. We perform a binned analysis in redshift to study if the $H_0$ tension also occurs in SNe Ia data. Hence, we build equally populated subsamples in three and four bins, and we estimate $H_{0}$ in each bin considering the $\Lambda$CDM and $w_{0}w_{a}$CDM cosmological models. We perform a statistical analysis via a Markov Chain Monte Carlo (MCMC) method for each bin. We observe that $H_0$ evolves with the redshift, using a fit function $H_{0}(z)=\tilde{H}_{0} (1+z)^{-\alpha}$ with two fitting parameters $\alpha$ and $\tilde{H}_{0}$. Our results show a decreasing behavior of $H_0$ with $\alpha\sim 10^{-2}$ and a consistency with no evolution between 1.2 $\sigma$ and 2.0 $\sigma$. Considering the $H_0$ tension, we extrapolate $H_{0}(z)$ until the redshift of the last scattering surface, $z=1100$, obtaining values of $H_0$ consistent in 1 $\sigma$ with the cosmic microwave background (CMB) measurements by Planck. Finally, we discuss possible $f(R)$ modified gravity models to explain a running Hubble constant with the redshift, and we infer the form of the scalar field potential in the dynamically equivalent Jordan frame. 

\end{abstract}

\maketitle

\section{Introduction \label{sec:intro}}

The Hubble constant tension is one of the biggest open problems of modern cosmology. Several independent measurements of the actual expansion rate of the Universe, the Hubble constant
$H_{0}$, provide incompatible values. The status of all these discrepancies and theoretical attempts
is summarized in a review \citep{reviewH0}. 
It should be noted that this inconsistency occurs between measurements referred to the early Universe, and those based on local probes (for instance, Cepheids or Type Ia supernovae - SNe Ia) in the late Universe. This tension has become more and more serious, little by little more precise measurements have reduced gradually the error bars. More precisely, there is a discrepancy in 4.4 $\sigma$ between the Planck data \citep{Planck2020}, $H_{0}=67.4\pm0.5\,\textrm{km s}^{-1}\,\textrm{Mpc}^{-1}$, and the value obtained from Cepheids \citep{Ceph}, $H_{0}=74.03\pm1.42\,\textrm{km s}^{-1}\,\textrm{Mpc}^{-1}$. This tension could point out inconsistencies in Planck data or local probes or, alternatively, it could be a signal for a new cosmology.

Regarding the latter point, modified gravity theories are studied to try to solve open problems in cosmology, such as the Hubble constant tension or the nature of dark energy, which cannot be fully explained in General Relativity (GR) and the standard $\Lambda$CDM cosmological model. In particular, the $f(R)$ modified gravity theories \citep{odintsov-f(R),sotiriou} provide a generalization of GR, including an extra geometrical degree of freedom given by a function $f$ of the Ricci scalar $R$, which implies a cosmological dynamics that differs from GR. The scalar-tensor theories in the so-called Jordan frame \citep{odintsov-f(R),sotiriou,book-capozz-faraoni} are dynamically equivalent to the $f(R)$ proposals; within this framework, the additional scalar degree of freedom is provided by a scalar field, which is non-minimally coupled to the metric.   

Since the $H_{0}$ tension concerns measurements referred to the early and late Universe, in this work we investigate the possibility of a hidden evolutionary effect that could imply a running Hubble constant with the redshift \citep{Krishnan2020,Krishnan2021,H0(z)1,H0(z)2}, focusing on the data analysis of the Pantheon sample \citep{scolnic}. In this regard, we follow a binned analysis of the Pantheon sample in three and four redshift bins to extract the values of $H_{0}$ in each bin. Then, we observe an unexpected decreasing trend of $H_{0}$ with the redshift, and we interpret this extra degree of freedom via a modified gravity scenario. 

This work is organized as follows: in Sec.~1 we recall briefly some basic notions about cosmology using SNe Ia datasets; in Sec.~2 we introduce the $f(R)$ extended models in the Jordan frame; in Sec.~3 we follow a binning approach to address the Hubble constant tension within the redshift range of the Pantheon sample; in Sec.~4 we discuss our results and the theoretical interpretations in the $f(R)$ modified cosmology.  

We adopt the metric signature $\left(-, +, +, +\right)$, and
we set the speed of light $c = 1$. The Einstein constant is denoted with $\chi\equiv8\,\pi\,G$, and $G$ is the Newton constant.

\section{Type Ia Supernova cosmology}
\label{sec:1}

SNe Ia are usually regarded as standard candles since they are characterized by their uniformity in the absolute magnitude profiles. Hence, they are very useful to estimate cosmic distances. In this regard, we recall the definition of the theoretical distance modulus of a SN: 
\begin{equation}
\mu_{\textrm{th}}=m_{th}-M=5\hspace{0.5ex}log_{10}\ d_{L}(z)+25\,,
\label{eq:mu-th}
\end{equation}
where $m_{th}$ and $M$ are the apparent and absolute magnitude of a SN, respectively, and $d_{L}$ is the luminosity distance, which depends on a specific cosmological model. In a flat geometry, $d_{L}$ is given by \citep{weinberg}
\begin{equation}
d_{L}\left(z\right)=\left(1+z\right)\,\int_{0}^{z}\frac{dz'}{H\left(z'\right)}\,.\label{eq:general form dl(z)}
\end{equation}
Considering a dark energy component in which the equation of state parameter $w$ evolves with the redshift $z$, i.e. $w=w(z)$, and neglecting relativistic components in the late Universe, the Hubble function $H(z)$ contained in Eq.~(\ref{eq:general form dl(z)}) is written as 
\begin{equation}
H\left(z\right)=H_{0}\,\sqrt{\Omega_{m0}\,\left(1+z\right)^{3}+\Omega_{DE\,0}\,\exp\left[3\,\int_{0}^{z}\left[1+w\left(z^{\prime}\right)\right]\frac{dz^{\prime}}{1+z^{\prime}}\right]}\,,\label{eq:H(z)_w(z)}
\end{equation}
where $\Omega_{m0}$, $\Omega_{DE\,0}\equiv 1-\Omega_{m0}$ are the cosmological density parameters of the matter and dark energy components, respectively, at the present redshift $z=0$. Note that if $w=-1$, a cosmological constant $\Lambda$ is reproduced, and the standard $\Lambda$CDM scenario is recovered.
Differently, considering in Eq.~(\ref{eq:H(z)_w(z)}) the $w_{0}w_{a}$CDM model, given by the Chevallier-Polarski-Linder (CPL) parameterization \citep{Chevallier,Linder} $w\left(z\right)=w_{0}+w_{a}\times z/\left(1+z\right)$ with parameters $w_0$ and $w_a$, we end up in a slowly evolving dark energy scenario slightly different from a cosmological constant.

Moreover, it should be noted that the Hubble constant $H_{0}$ is degenerate with $M$, as you can see easily combining Eqs.~\eqref{eq:mu-th} \eqref{eq:general form dl(z)} and \eqref{eq:H(z)_w(z)}. Specifically, $M$ is calibrated to $-19.35$ in the Pantheon sample such that $H_{0}=70.0\,\textrm{km s}^{-1}\,\textrm{Mpc}^{-1}$ \citep{scolnic}.

To discriminate between several cosmological models, it is useful
to compare $\mu_{\text{th}}$ in Eq.~\eqref{eq:mu-th} with the observed distance modulus $\mu_{\text{obs}}$, which is obtained from SNe Ia data, 
\begin{equation}
\mu_{\textrm{obs}}=m_{B}-M+\alpha\,x_{1}-\beta\,c+\Delta M+\Delta B\,,\label{eq:mu-obs}
\end{equation}
where $m_{B}$ is the $\textit{B}$-band apparent magnitude, and $M$ is the absolute magnitude in the $\textit{B}$-band$x_{1}$ of a SN with $c=0$ and $x_{1}=0$. SNe parameters are the color $c$ and the stretch $x_1$, while $\alpha$ and $\beta$ are coefficients. Finally, $\Delta M$ is a distance correction related to the host-galaxy mass, and $\Delta B$ is a bias correction. The Pantheon sample data is available in the repository by Scolnic et al. (2018) \citep{scolnic}  (\href{https://github.com/dscolnic/Pantheon}{https://github.com/dscolnic/Pantheon}). 

To perform a statistical analysis, we define $\Delta\mu=\mu_{\textrm{obs}}-\mu_{\textrm{th}}$ as the difference between the distance moduli in Eqs.~\eqref{eq:mu-th} and \eqref{eq:mu-obs}. Furthermore, the $\chi^{2}$ is built as  
\begin{equation}
\chi^{2}=\Delta\mu^{T}\,C^{-1}\,\Delta\mu\,,
\label{eq:chi2}
\end{equation}
where $C=C_{sys}+D_{stat}$ is the full covariance matrix. The $C_{sys}$ matrix involves systematic errors, while $D_{stat}$ is a diagonal matrix that contains statistical errors for each SN, due to peculiar velocities, photometry, bias, lensing, and intrinsic scatter \citep{scolnic}.
Referring to the Pantheon sample, $\Delta\mu$ represents a vector
with $1048$ components, while $C$,
$C_{sys}$ and $D_{stat}$ are $1048\times1048$ square matrices.

\section{f(R) modified cosmology in the Jordan frame}

\label{sec:intro-modified-gravity}

Within the framework of the $f(R)$ modified gravity \citep{odintsov-f(R),sotiriou}, the gravitational Lagrangian density becomes a function $f(R)$ of the scalar curvature $R$, involving an additional scalar degree of freedom with respect to GR. Adopting the metric formalism, it can be noted that the generalized field equations are of fourth-order in the metric \citep{sotiriou}, and if $f\left(R\right)=R$ one might obtain the Einstein-Hilbert equations in GR. 
The scalar-tensor theories in the so-called Jordan frame \citep{odintsov-f(R),sotiriou,book-capozz-faraoni} are useful to rewrite $f(R)$ models in a dynamically equivalent form, which provides field equations of lower order. 
The total action in the Jordan frame is written as
\begin{equation}
S_{J}=\frac{1}{2\,\chi}\,\int d^{4}x\,\sqrt{-g}\,\left[\phi\,R-V\left(\phi\right)\right]+S_{M}\left(g_{\mu\nu},\psi\right)\,,
\label{eq: azione jordan frame}
\end{equation}
where $g$ is the determinant of $g_{\mu\nu}$ the metric tensor, $S_M$ is the matter contribution to the total action, and $\psi$ is referred to the matter fields. Moreover, in this paradigm, the extra degree of freedom given by $f\left(R\right)$ is expressed as a scalar field $\phi$, defined as $\phi=f^{\prime}\left(R\right)=df/dR$, which is non-minimally coupled to the metric, and it is governed by the scalar field potential $V\left(\phi\right)=\phi\,R\left(\phi\right)-f\left(R\left(\phi\right)\right)$.

The field equations in the Jordan frame for a flat FLRW geometry in the late Universe, considering a pressureless matter component and neglecting relativistic species, are given by the following equation system \citep{book-capozz-faraoni}:
\begin{subequations}
\begin{align}
& H^{2} =\frac{\chi\,\rho}{3\,\phi}-H\,\frac{\dot{\phi}}{\phi}+\frac{V\left(\phi\right)}{6\,\phi}\label{eq:generalized-Friedmann}\\
& \frac{\ddot{a}}{a} =-\frac{\chi\,\rho}{6\,\phi}+\frac{V\left(\phi\right)}{6\,\phi}-\frac{H}{2}\,\frac{\dot{\phi}}{\phi}-\frac{1}{2}\,\frac{\ddot{\phi}}{\phi}\label{eq:generalized-acc}\\
& 3 \ddot{\phi}-2\,V\left(\phi\right)+\phi\,\frac{dV}{d\phi}+9\,H\,\dot{\phi}=\chi\,\rho,\label{eq:scalar-fieldFLRW}
\end{align}
\end{subequations}
where $\rho\left(t\right)$ is the matter energy density, and a dot denotes a time derivative. The first equation above is the modified Friedmann equation, the second one is the modified acceleration equation, while the third equation describes the scalar field dynamics. Note that all these equations are now of second order in the equivalent formalism in the Jordan frame, but the disadvantage is the introduction of a non-minimally coupling between the scalar field and the metric.

It should be emphasized that the gravitational coupling constant is restated in the Jordan frame. More specifically, looking at Eq.~\eqref{eq:generalized-Friedmann}, $G_{\text{eff}}=G / \phi$ may be regarded as an effective gravitational coupling. 
Modified gravity models are very interesting to explore alternative scenarios for the cosmic acceleration in the late Universe without a true cosmological constant (see the models by \citep{sotiriou,hu-sawicki,starob,tsujik}).

Finally, following the Hu-Sawicki formalism \citep{hu-sawicki}, we introduce auxiliary variables $y_H$ and $y_R$ that are useful to write the luminosity distance in the $f(R)$ gravity. We write the Hubble parameter and Ricci scalar as
\begin{equation}
    H^2=m^2 \left(\left(1+z\right)^3 +  y_H\right),\hspace{10ex}
    R=m^2 \left(3\left(1+z\right)^3 +  y_R\right),
    \label{eq_yH_yR}
\end{equation}
where the two dimensionless variables $y_H$ and $y_R$ encompass extra contributions with respect to the matter component in the $\Lambda$CDM model. We have defined $m^{2}=\chi\,\rho_{0}/3=H_{0}^{2}/\Omega_{m0}$ with $\rho_0$ present matter density. Note that if $y_H$ is simply a constant, the first relation is nothing more the Friedmann equation in the $\Lambda$CDM model. In the context of $f(R)$ gravity, instead, $y_H$ and $y_R$ evolve and their dynamics strongly depend on the form of the $f(R)$ function or, equivalently, the scalar field potential $V\left(\phi\right)$. It is possible to rewrite the generalized Friedmann equation \eqref{eq:generalized-Friedmann} and the Ricci scalar in terms of $y_H$ and $y_R$ and their derivatives to obtain a fist-order differential equation system (see \citep{hu-sawicki}), which can be solved numerically. 
Then, we can write the luminosity distance in a $f(R)$ model
 \begin{equation}
    d_L(z)=\frac{(1+z)}{H_0}\int^{z}_{0} \frac{dz'}{\sqrt{\Omega_{m0}\biggr(y_H(z') + (1+z')^3\biggr)}}
    \label{eq_distlumin_new}
 \end{equation}
in terms of the deviation $y_H(z)$, combining Eqs.~\eqref{eq:general form dl(z)} and \eqref{eq_yH_yR}. 
 
\section{Binned analysis}
\label{sec:binning}

We decide to follow a binned approach using the Pantheon sample dataset to evaluate if the Hubble constant evolves with the redshift \citep{H0(z)1,H0(z)2}, motivated by the fact the independent measurements of $H_0$ referred to different redshifts had shown a tension. 

Therefore, we have split the redshift range of the Pantheon sample in bins each having the same number $N$ of SNe Ia. More precisely, considering three redshift bins, we have $N\approx 349$ SNe for each bin, and the redshift ranges for three bins are $0.01<z<0.18$,
$0.18<z<0.34$, and $0.34<z<2.26$. In the case of four bins, instead, $N=262$ in each bin and the redshift ranges are: $0.01<z<0.13$, $0.13<z<0.25$, $0.25<z<0.42$, and $0.42<z<2.26$.
Then, we have built the subvectors $\Delta\mu$ with $N$ SNe, according to the binning division and redshift order. We have also divided the full covariance matrix $C$ into $N\times N$ submatrices for each redshift bin. In this regard, concerning simply the statistical contribution $D_{stat}$ in $C$, each diagonal element of the $D_{stat}$ matrix is related to a single SN, hence we can quickly build the submatrices starting from $D_{stat}$. However, if we include the systematic errors, we need to consider also $C_{sys}$, which is not a diagonal matrix. Thus, we used a customized code to select $C_{sys}$ elements involved only with the SNe within the redshift bin taken into account. Hence, we built properly the submatrices, considering both statistical and systematic errors related to the SNe for each bin. 

To focus only on $H_{0}$ in a one-dimensional analysis and constrain it in each bin, we decide to fix the value of the other cosmological parameters. Thus, we assume the fiducial value \citep{scolnic} $\Omega_{m0}=0.298$ for a flat $\Lambda$CDM model, and $\Omega_{m0}=0.308$ with $w_{0}=-1.009$ and $w_{a}=-0.129$ for a flat $w_{0}w_{a}$CDM model for all the redshift ranges. Moreover, we need to fix the absolute magnitude $M$ of SNe Ia in each bin. Bearing in mind the degeneracy between $M$ and $H_{0}$, we decide to calibrate $M$ to obtain a value of $H_{0}=73.5\,\textrm{km s}^{-1}\,\textrm{Mpc}^{-1}$ in the first redshift bins, according to measurements of the local probes at low redshifts.

Thus, we perform a preliminary analysis for both $\Lambda$CDM and $w_{0}w_{a}$CDM models to obtain the respective value of $M$
for each binning division in the first redshift bin, which is characterized by a mean redshift $z\ll1$. 
To do this, we minimize the $\chi^{2}$ in Eq.~\eqref{eq:chi2}. Then, we use the MCMC methods to sample a posterior distribution. Finally, $M$ is obtained for each binning division (see Table~\ref{TableH073}). After this preliminary analysis, we fix $M$ in the other bins. One might expect that also $H_{0}$ assumes the same values in all the redshift bins, due to its degeneracy with $M$. 

\begin{table} \begin{centering} \begin{tabular}{|c|c|c|c|c|c|c|} \hline \multicolumn{7}{|c|}{Flat $\Lambda$CDM Model, Fixed $\Omega_{m0}=0.298$}\tabularnewline 
\hline Bins & $\tilde{H}_0$ & $\alpha$ & $\frac{\alpha}{\sigma_{\alpha}}$ & $M$ & $H_{0}\left(z=11.09\right)$ & $H_{0}\left(z=1100\right)$ \\ & $\,\left(\textrm{km s}^{-1}\,\textrm{Mpc}^{-1}\right)$ & & & & $\,\left(\textrm{km s}^{-1}\,\textrm{Mpc}^{-1}\right)$ & $\,\left(\textrm{km s}^{-1}\,\textrm{Mpc}^{-1}\right)$ \tabularnewline 
\hline 3 & $73.577\pm0.106$ & $0.009\pm0.004$ & $2.0$ & $-19.245\pm0.006$ & $72.000\pm0.805$ & $69.219\pm2.159$ \tabularnewline 
\hline 4 & $73.493\pm0.144$ & $0.008\pm0.006$ & $1.5$ & $-19.246\pm0.008$ & $71.962\pm1.049$ & $69.271\pm2.815$ \tabularnewline 
\hline \hline \multicolumn{7}{|c|}{Flat $w_{0}w_{a}$CDM Model, Fixed $\Omega_{m0}=0.308$, $w_0=-1.009$ and $w_{a}=-0.129$ }\tabularnewline 
\hline Bins & $\tilde{H}_0$ & $\alpha$ & $\frac{\alpha}{\sigma_{\alpha}}$ & $M$ & $H_{0}\left(z=11.09\right)$ & $H_{0}\left(z=1100\right)$ \\ & $\,\left(\textrm{km s}^{-1}\,\textrm{Mpc}^{-1}\right)$ & & & & $\,\left(\textrm{km s}^{-1}\,\textrm{Mpc}^{-1}\right)$ & $\,\left(\textrm{km s}^{-1}\,\textrm{Mpc}^{-1}\right)$ \tabularnewline 
\hline 3 & $73.576\pm0.105$ & $0.008\pm0.004$ & $1.9$ & $-19.244\pm0.005$ & $72.104\pm0.766$ & $69.516\pm2.060$ \tabularnewline 
\hline 4 & $73.513\pm0.142$ & $0.008\pm0.006$ & $1.2$ & $-19.246\pm0.004$ & $71.975\pm1.020$ & $69.272\pm2.737$ \tabularnewline 
\hline \end{tabular} 
\caption{Fitting parameters ($\tilde{H}_0$ and $\alpha$) of $H_0(z)$ (Eq.~(\ref{eq:H0(z)}) and extrapolated values at high redshifts, after a binned analysis of the Pantheon sample, focusing on a flat $\Lambda$CDM model (upper part) and a flat $w_{0}w_{a}$CDM model (lower part). The compatibility of $\alpha$ with zero (no evolution) is expressed in terms of 1 $\sigma$ in the fourth column. The new absolute magnitude $M$, which provides $H_0=73.5\, \textrm{km s}^{-1}\,\textrm{Mpc}^{-1}$ in the first bins, is listed in the fifth column. Finally, the extrapolated values of $H_{0}(z)$ at the redshift of the most distant galaxies, $z=11.09$, and the last scattering surface, $z=1100$, are listed in the last two columns. All the errors are presented in 1 $\sigma$.} \label{TableH073} \par\end{centering} \end{table}

\begin{figure} \includegraphics[scale=0.125]{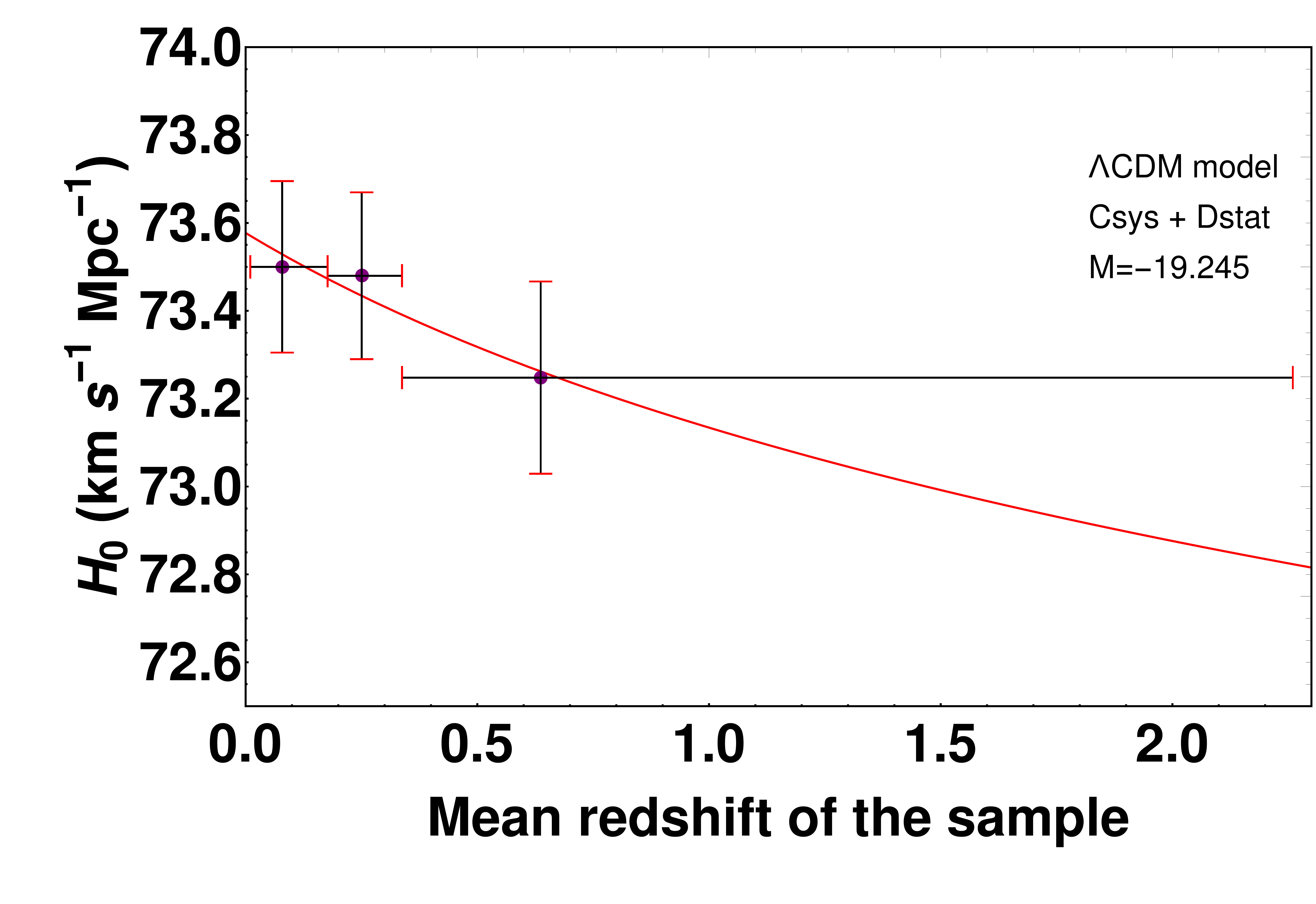} \includegraphics[scale=0.125]{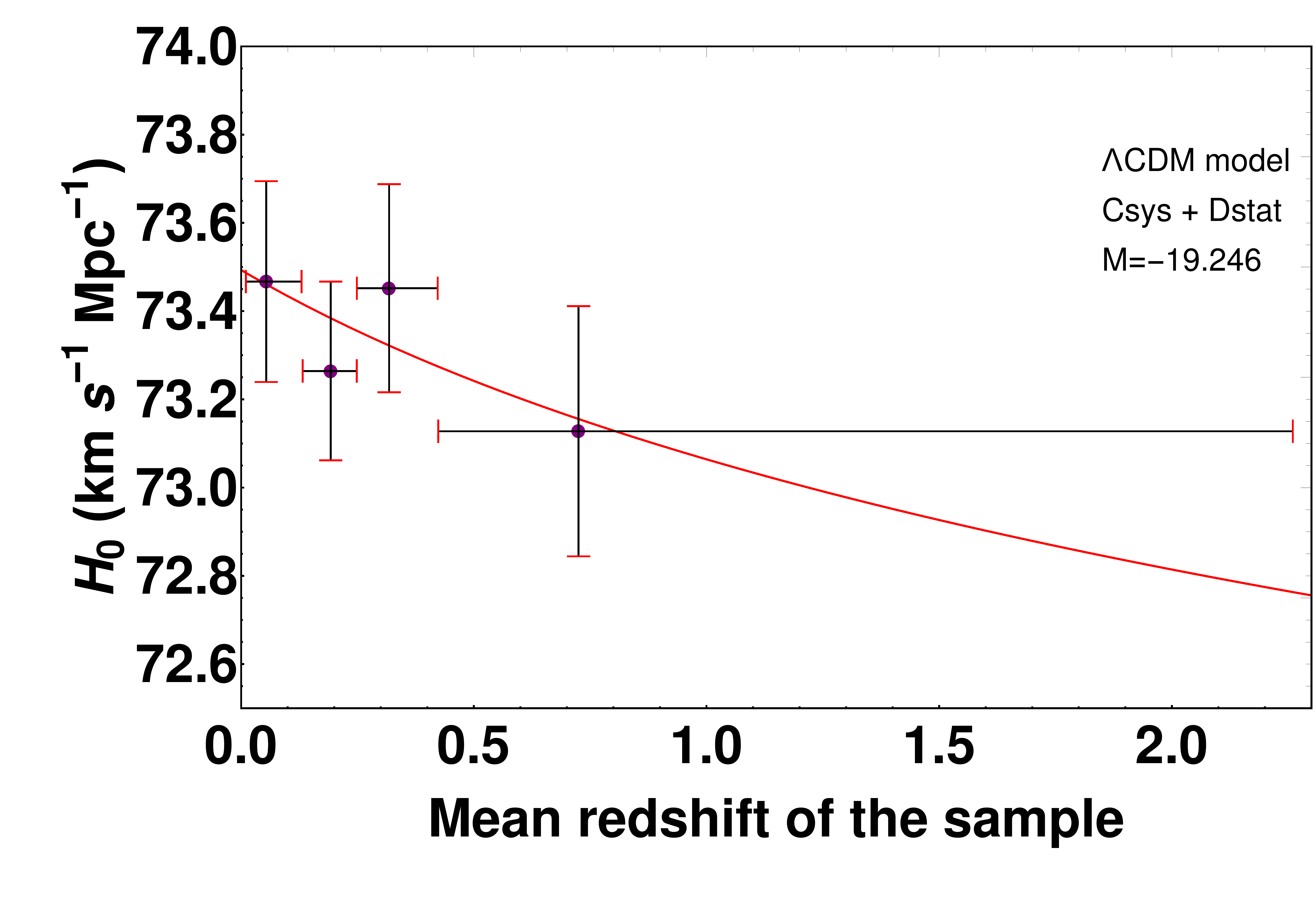} \\ \includegraphics[scale=0.125]{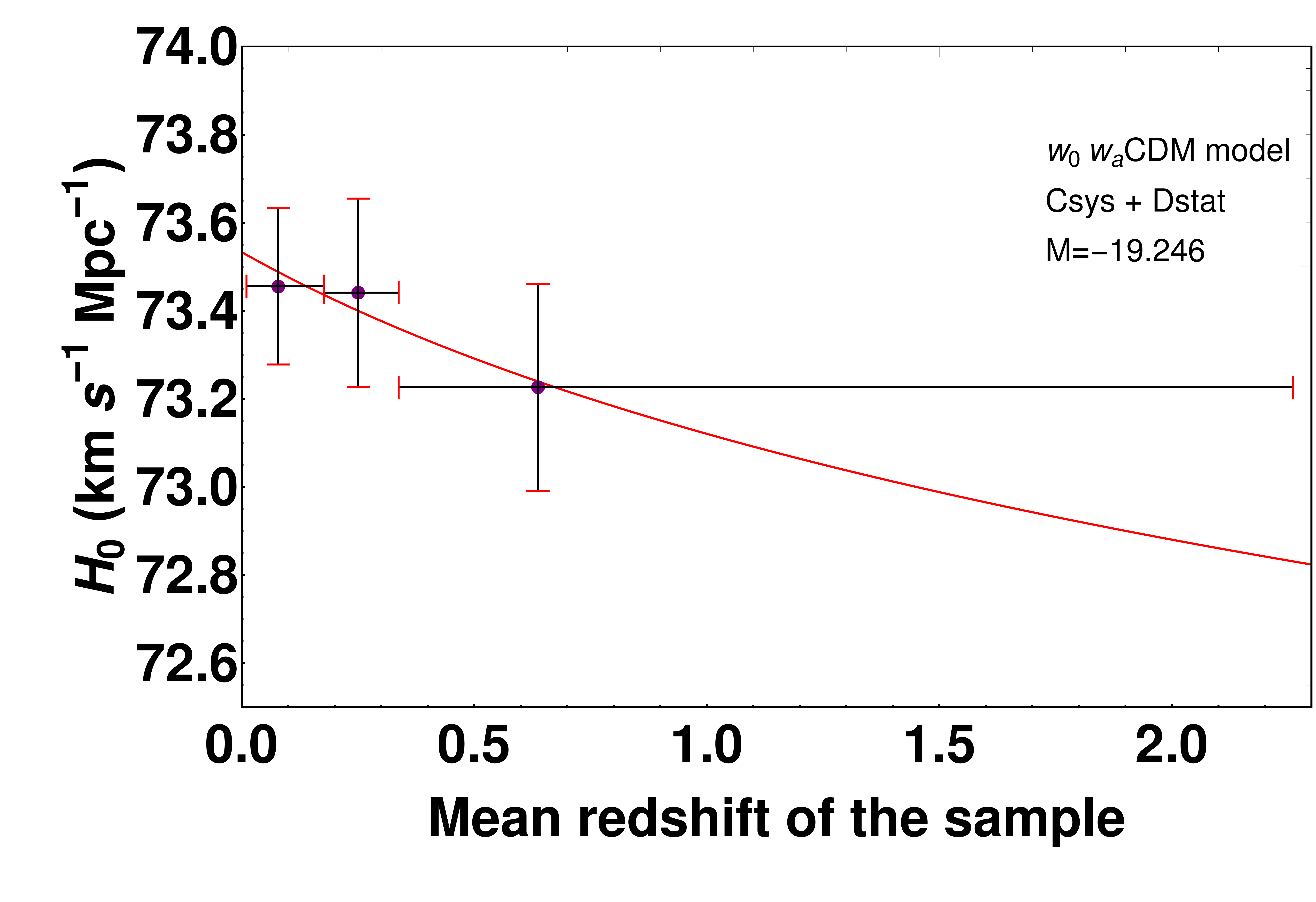} \includegraphics[scale=0.125]{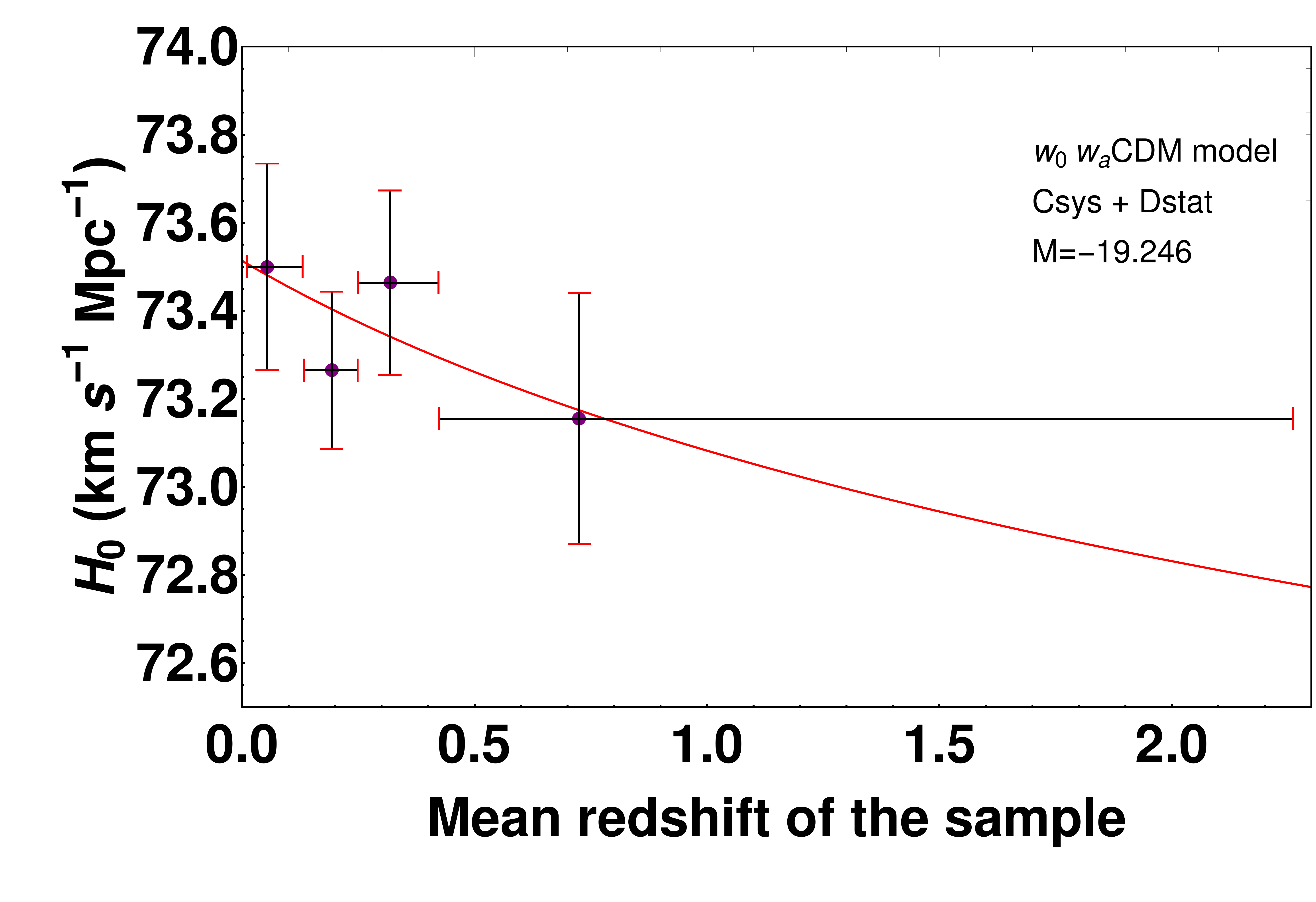} \caption{Evolution of $H_0(z)$ with redshift according to Eq.~\eqref{eq:H0(z)} as a result of a binned analysis of the Pantheon sample for a flat $\Lambda$CDM model (two upper plots) and a flat $w_{0}w_{a}$CDM model (two lower plots).} \label{figH073} \end{figure}

Now, we can approach the main part to check the values of $H_0$ after a binned analysis of the Pantheon Sample in three and four bins for both the $\Lambda$CDM and $w_{0}w_{a}$CDM
models.   
We use again the minimization of $\chi^{2}$, as illustrated in the preliminary analysis, and in the MCMC we set the priors: $60$ $\textrm{km s}^{-1}\,\textrm{Mpc}^{-1}$ $<H_{0}<80$ $\textrm{km s}^{-1}\,\textrm{Mpc}^{-1}$. Finally, we have extracted the values of $H_{0}$ for the all bins. To investigate an evolution of $H_0$ with the redshift $z$, we apply a non-linear
fit \citep{H0(z)1} of $H_{0}(z)$ written as 
\begin{equation}
H_{0}(z)=\frac{\tilde{H}_{0}}{(1+z)^{\alpha}}\,,
\label{eq:H0(z)}
\end{equation}
with two fitting parameters, $\tilde{H}_{0}$ and $\alpha$. Note
that $\tilde{H}_{0}=H_{0}\left(z=0\right)$. Moreover,
if $\alpha\neq0$, we have an evolutionary trend, otherwise $H_{0}$
is a constant. 

The fitting parameters and their respective errors in 1 $\sigma$ are listed in Table~\ref{TableH073}. Note that, for instance, the $\alpha$ parameter is consistent with zero, namely no evolution, in 2.0 $\sigma$ in the $\Lambda$CDM model using three bins.
In Fig.~\ref{figH073} we can observe a slowly and unexpected decreasing trend \citep{H0(z)1} for $H_{0}\left(z\right)$ in the $\Lambda$CDM model and also in the $w_{0}w_{a}$CDM model. It should be noted that all the $\alpha$ coefficients are mutually compatible in 1 $\sigma$, therefore pointing out a reliable decreasing trend of $H_0$ with the redshift in the Pantheon sample. 

If the observed trend of $H_{0}(z)$ is intrinsic and does not depend on a specific sample, one may ask what happens with other local probes or in the early Universe at very high redshifts. Hence, we extrapolate the fit function of $H_{0}(z)$ at the redshift of the most distant galaxies, $z=11.09$
\citep{oesch}, and at the redshift of the last scattering surface, $z=1100$, to compare the latter values of $H_{0}(z)$ with the one inferred from Planck measurements for the CMB.
We find that the extrapolated values (see Table~\ref{TableH073}) are consistent within 1 $\sigma$ with the value of
$H_{0}$ obtained with the Planck measurements for both the $\Lambda$CDM and $w_{0}w_{a}$CDM models regardless the number of bins.

\section{Interpretation of the results in the f(R) modified cosmology}

\label{sec:Theoretical-discussions}

The evolution of the Hubble constant with the redshift needs a physical interpretation. In this section, we do not focus on possible astrophysical reasons (see \citep{H0(z)1}), but we investigate our results in modified gravity theories. 

The Hubble constant must be a constant by definition, and its observed evolution could be a signal of a wrong framework, for instance, a possible hidden function of the redshift, which has not been taken into account so far. 
As already discussed in \citep{Kaz,H0(z)1}, a varying Einstein constant $\chi=8\pi G$ (or a varying Newton constant $G$) may in principle lead to an evolution of $H_{0}\left(z\right)$. If we look at the function of $H_{0}\left(z\right)$ in Eq.~(\ref{eq:H0(z)}), we require an effective Einstein constant $\chi\sim\left(1+z\right)^{-2\,\alpha}$ to preserve constant the present critical density $\rho_{c0}=3\,H_{0}^{2}/\chi$.

We recall that an effective Einstein constant is naturally obtained in the Jordan frame (Sec.~2) through the rescaling: $\chi\rightarrow\chi\,/\,\phi$. The extra degree of freedom in the Jordan frame, i.e. the presence of a non-minimally coupled scalar field $\phi$, which depends on the redshift, provides indeed an evolving Einstein constant. Moreover, another reason to focus on the Jordan frame to account for the observed trend of $H_{0}(z)$ lies in the modified Friedmann equation \eqref{eq:generalized-Friedmann}. Indeed, the Hubble parameter $H(z)$ is related to matter density $\rho$ via an effective Einstein constant, hence the dynamics of the scalar field $\phi$ may imply an evolution of the Hubble constant with redshift. More specifically, we require the following ansatz
\begin{equation}
\phi\left(z\right)=\left(1+z\right)^{2\,\alpha}
\label{eq:phi(z)}
\end{equation}
to account for $H_{0}\left(z\right)$ in Eq.~(\ref{eq:H0(z)}).  
These concepts suggest that an extra degree of freedom with respect to GR might imply the unexpected trend of $H_{0}\left(z\right)$. 

We do not mention so far which scalar field potential we should have in the Jordan frame, mimicking a cosmological constant in a slow-roll and at the same time providing an effective Hubble constant. Considering the cosmological dynamics in the Jordan frame, we can infer $V\left(\phi\right)$ \citep{H0(z)2}, if we assume the behavior of $H_{0}\left(z\right)$ in Eq.~(\ref{eq:H0(z)}) and the ansatz above for $\phi\left(z\right)$ in Eq.~\eqref{eq:phi(z)}.  
Therefore, the modified Friedmann equation (\ref{eq:generalized-Friedmann}) allows us to write $V\left(\phi\right)$ as 
\begin{equation}
V\left(\phi\right)=6\,(1-2\alpha)\,\left(\frac{dz}{dt}\right)^2\,\phi^{1-1/\alpha}-6 m^{2}\,\phi^{3/2\alpha}\,,
\label{eq:V(phi)dz/dt}
\end{equation}
where we have used the standard definition of redshift, and $\rho\sim (1+z)^3$ for a matter component. Now, we need to specify the term $dz/dt$, which is given by
\begin{equation}
\frac{dz}{dt}=-(1+z)\,H(z).
\label{eq:dz/dt}
\end{equation}
Here, we can not simply use the extended Friedmann equation (\ref{eq:generalized-Friedmann}) to replace $H(z)$, because we should set a specific scalar field potential to solve $H(z)$. Then, we decide to impose the Hubble parameter 
\begin{equation}
H\left(z\right)=\frac{\tilde{H}_{0}}{(1+z)^\alpha}\sqrt{\Omega_{m0}\,\left(1+z\right)^{3}+1-\Omega_{m0}} \,,
\label{eq:H(z)-inferred}
\end{equation}
as suggested from our binned analysis in Sec.~3, although its analytical form is different from the Hubble parameter obtained using the modified Friedmann equation. In other words, to obtain $V\left(\phi\right)$, we impose in the Jordan frame dynamics the same physical effect observed from the redshift binned analysis of the Pantheon sample. 

\begin{figure}
    \centering
    \includegraphics[scale=0.18]{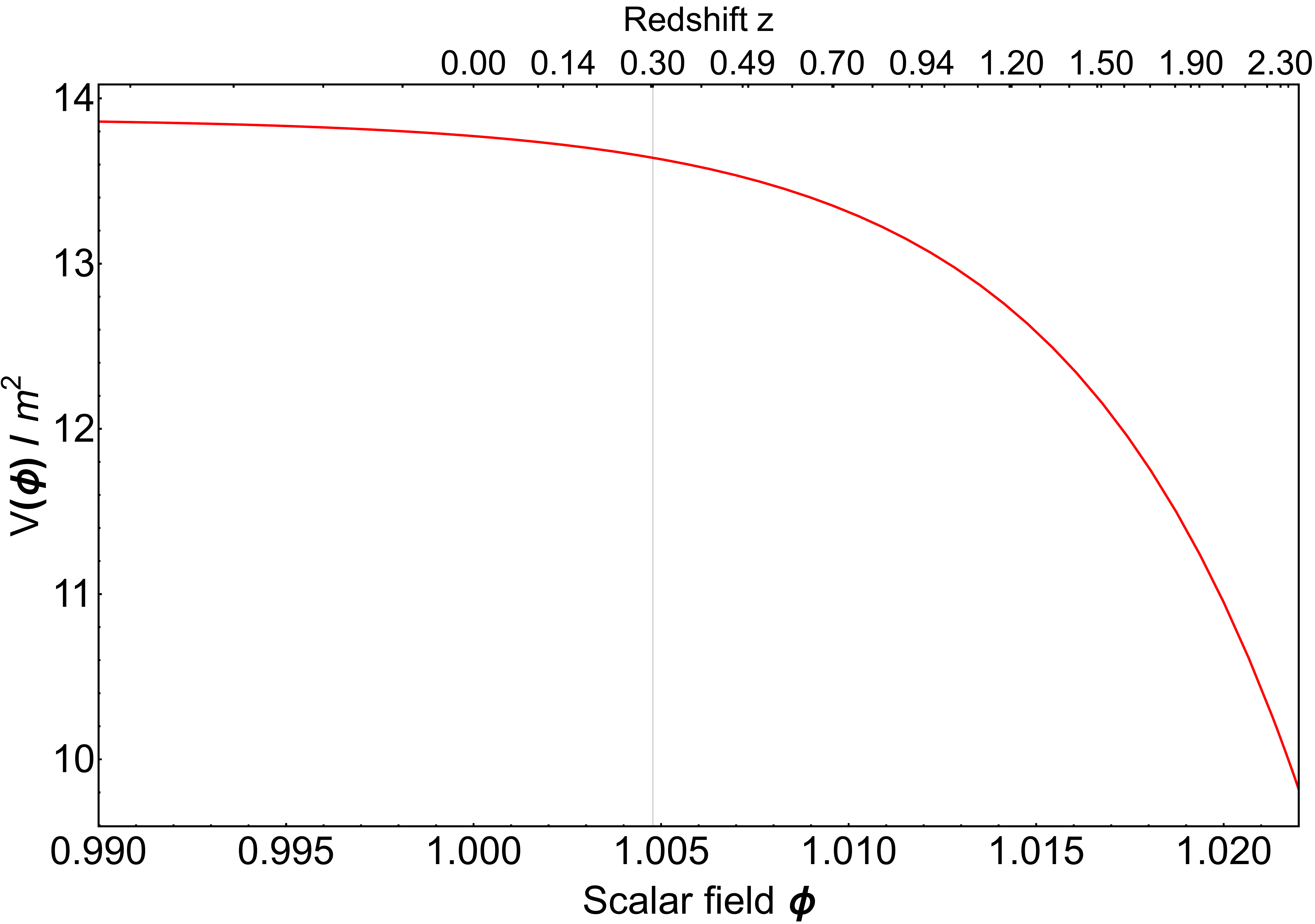}
    \caption{Scalar field potential $V(\phi)$ in the Jordan frame given by Eq.~\eqref{eq:final-potential} and inferred from the evolution of $H_{0}\left(z\right)$ with redshift (\ref{eq:H0(z)}). The quantity $V(\phi)/m^2$ is dimensionless. Note the presence of a flat region of the potential for $0<z\lesssim 0.3$ or $\phi\lesssim 1.005$. }
    \label{fig:scalar-field}
\end{figure}

Hence, combining Eqs.~\eqref{eq:V(phi)dz/dt}, \eqref{eq:dz/dt} and \eqref{eq:H(z)-inferred}, we obtain the final expression of the potential \citep{H0(z)2}:
\begin{equation}
\frac{V\left(\phi\right)}{m^2}=6\,(1-2\alpha)\,\frac{1-\Omega_{m0}}{\Omega_{m0}}-12\,\alpha\,\phi^{\frac{3}{2\alpha}}\,,
\label{eq:final-potential}
\end{equation}
where we have considered the relation $\Omega_{m0}=m^{2}/\tilde{H}_{0}^2$.
You can see the form of the scalar field potential in Fig.~\ref{fig:scalar-field}, where we fixed $\alpha=0.009$ (see Table~\ref{TableH073}) and $\Omega_{m0}=0.298$ for plotting $V\left(\phi\right)/m^2$. Note the occurrence of a flat region of $V\left(\phi\right)$ for $0<z\lesssim 0.3$, mimicking a cosmological constant in the era dominated by dark energy, thus validating our approach and the inferred scalar field potential.

Furthermore, we can obtain the $f(R)$ function related to $V\left(\phi\right)$ in Eq.~\eqref{eq:final-potential}. Using the field equation $R=dV/d\phi$ in the Jordan frame, and also the relation $f(R)=R\,\phi(R)-V\left(\phi(R)\right)$ \citep{sotiriou}, we write:
\begin{equation}
f\left(R\right)=-6\,m^{2}\,\left[\left(3-2\alpha\right)\,\left(-\frac{R}{18\,m^2}\right)^{\frac{3}{3-2\alpha}}+\left(1-2\alpha\right)\frac{1-\Omega_{0m}}{\Omega_{0m}}\right]\,.
\label{eq:total-f(R)}
\end{equation}
Expanding this expression for $\alpha\sim0$, according to the values of $\alpha$ from the binned analysis (see Table~\ref{TableH073}), we obtain \citep{H0(z)2}:
\begin{equation}
f\left(R\right)\approx \left(R-6\,m^{2}\,\frac{1-\Omega_{0m}}{\Omega_{0m}}\right)+\frac{2}{3}\alpha\,\left[R\,\ln{\left(-\frac{R}{m^2}\right)}-\left(1+\ln{18}\right)\,R+18m^{2}\,\frac{1-\Omega_{0m}}{\Omega_{0m}}\right]+O\left(\alpha^2\right)\,.
\label{eq:f(R)-expansion}
\end{equation}
It should be emphasized that the first term above is precisely the gravitational Lagrangian density in GR with a cosmological constant $\Lambda=3m^{2}\left(1-\Omega_{0m}\right)\,/\,\Omega_{0m}$, where we used $m^{2}=\tilde{H}^{2}_{0}\,\Omega_{0m}$. Moreover, it is clear that the first-order term in $\alpha$ in Eq.~\eqref{eq:f(R)-expansion} gives the deviation from GR. Hence, the $\alpha$ coefficient stresses corrections to GR.  

The expressions inferred for the scalar field potential in Eq.~\eqref{eq:final-potential}, and the corresponding $f(R)$ function in Eq.~\eqref{eq:f(R)-expansion}, are viable in the late Universe since in all the calculations we do not include relativistic components. However, it could be interesting to test this form of $V\left(\phi\right)$ using other probes in the late Universe. 

The proposed discussions above show a possible interpretation of our results coming from the binned analysis, and we suggest a candidate for $V\left(\phi\right)$, assuming the evolution of $H_0(z)$. 

\section{Conclusions}
In this work, performing a redshift binned analysis of the Pantheon sample of SNe Ia, we have seen that there is a slow evolution of the Hubble constant $H_{0}$ with the redshift $z$, described by Eq.~\eqref{eq:H0(z)} with a parameter $\alpha$ that is consistent with zero (no evolution) between 1.2 $\sigma$ and 2.0 $\sigma$. 
The evolutionary behavior of $H_{0}(z)$ occurs regardless of the binning division both in the $\Lambda$CDM and $w_{0}w_{a}$CDM models. It seems that the Hubble constant tension also emerges locally in the narrow redshift range $0<z<2.26$ of the Pantheon sample of SNe Ia. 

Then, extrapolating the fit function of $H_{0}(z)$ (\ref{eq:H0(z)}) at higher redshifts, we obtain values of $H_{0}$ that are remarkably consistent within 1 $\sigma$ with the CMB measurements by Planck. 
Our results could point out an intrinsic trend of $H_{0}(z)$, which might imply the mismatch between independent measurements of the Hubble constant referred to the early and late Universe.

We remark that we perform a one-dimensional analysis to obtain constraints only on $H_0$, following \citep{H0(z)1}. Actually, in \citep{H0(z)2} we extend the analysis, and we obtain similar results, considering more than one variable in the MCMC methods, for instance $H_0$ and $\Omega_{m0}$, to avoid fixing all the cosmological parameters except $H_0$.

Our analysis could suggest the presence of a hidden astrophysical effect, which is the reason for the evolution of $H_{0}(z)$ and has not been taken into account so far. Alternatively, the observed effect may point out that we need a cosmology beyond the $\Lambda$CDM paradigm. We have discussed the $f(R)$ modified gravity theories, and we suggest a possible form for the scalar field potential (\ref{eq:final-potential}) in the Jordan frame, and the respective $f(R)$ function (\ref{eq:f(R)-expansion}).
Furthermore, we mention that a new binned analysis of the Pantheon sample could be worthwhile, using a specific $f(R)$ model and the modified version of the luminosity distance (\ref{eq_distlumin_new}) to test new physics.

Concerning the future perspectives, to explore further the possible evolution of cosmological parameters, it could be very useful to include in our analysis also probes at high redshifts, such as the Gamma-Ray Bursts (GRBs), which in principle could allow to enlarge the redshift range of the actual Hubble diagram \citep{14Cardone2009,15Cardone2010,17Dainotti2013,18Postnikov2014,361Dainotti2021,362Dainotti2019}. This work could be regarded as a preliminary study for the application of GRBs in cosmology, using, for instance, the prompt-afterglow relations with the plateau emission \citep{24Dainotti2008,25Dainotti2010,26Dainotti2011,27Dainotti2013,28Dainotti2015,29Dainotti2015,30Dainotti2016,31Dainotti2017,32Dainotti2017,33Dainotti2018,34Dainotti2018,35Dainotti2020,36Dainotti2021,37Dainotti2021,38DelVecchio2016,349Dainotti2021,353Cao2022}, similarly to the magnetar model \citep{41Rowlinson2014,42Rea2015,43Stratta2018}.

In conclusion, we have presented a new perspective to address the Hubble constant tension through a binning approach, thus we foster investigations on the astrophysical parameters of SNe Ia and further theoretical discussions.

\nocite{*}

\begin{thebibliography}{99}

\bibitem{reviewH0}
E. Di Valentino, O. Mena, P. Supriya, {\it et al.}, In
the realm of the Hubble tension --- a review of solutions, \href{https://iopscience.iop.org/article/10.1088/1361-6382/ac086d}{{\em Class. Quantum Grav.} {\bf 38}, 153001} (2021)

\bibitem{Planck2020}
N. Aghanim, Y. Akrami, M. Ashdown, {\it et al.}, Planck
2018 results. VI. Cosmological parameters, \href{https://www.aanda.org/articles/aa/full_html/2020/09/aa33910-18/aa33910-18.html}{{\em A$\&$A} {\bf 641}, A6} (2020)

\bibitem{Ceph}
M. J. Reid, D. W. Pesce and A. G. Riess, An Improved Distance to NGC 4258 and Its Implications for the Hubble Constant, \href{https://ui.adsabs.harvard.edu/abs/2019ApJ...886L..27R}{{\em ApJL} {\bf 886}, L27} (2019)

\bibitem{odintsov-f(R)}
S. Nojiri, and S. D. Odintsov, Introduction
to modified gravity and gravitational alternative for dark energy,
\href{https://www.worldscientific.com/doi/abs/10.1142/S0219887807001928}{{\em IJGMM} {\bf 4}, 115} (2007)

\bibitem{sotiriou}
T. P. Sotiriou, and V. Faraoni, $f\left(R\right)$
theories of gravity, \href{https://journals.aps.org/rmp/abstract/10.1103/RevModPhys.82.451}{{\em Rev. Mod. Phys.} {\bf 82}, 451} (2010)

\bibitem{book-capozz-faraoni}
S. Capozziello, and V. Faraoni, {\em Beyond
Einstein gravity} (Springer, 2013)

\bibitem{Krishnan2020}
C. Krishnan, E. Ó Colgáin, Ruchika, {\it et al.}, Is there an early Universe solution to Hubble tension?, \href{https://journals.aps.org/prd/abstract/10.1103/PhysRevD.102.103525}{{\em Phys. Rev. D} {\bf 102}, 103525} (2020)

\bibitem{Krishnan2021}
C. Krishnan, E. Ó Colgáin, M.M. Sheikh-Jabbari, and Tao Yang, Running Hubble tension and a H0 diagnostic, \href{https://journals.aps.org/prd/abstract/10.1103/PhysRevD.103.103509}{{\em Phys. Rev. D} {\bf 103}, 103509} (2021)

\bibitem{H0(z)1}
M. G. Dainotti, B. De Simone, T. Schiavone, {\it et al.}, On the Hubble Constant Tension in the SNe Ia Pantheon Sample, \href{https://iopscience.iop.org/article/10.3847/1538-4357/abeb73}{{\em ApJ} {\bf 912(2)}, 150} (2021)

\bibitem{H0(z)2}
M. G. Dainotti, B. De Simone, T. Schiavone, {\it et al.}, On the Evolution of the Hubble constant with the SNe Ia Pantheon Sample and Baryon Acoustic Oscillations: a Feasibility Study for GRB-Cosmology in 2030, \href{https://www.mdpi.com/2075-4434/10/1/24}{{\em Galaxies} {\bf 10}, 24} (2022) 

\bibitem{scolnic}
D. M. Scolnic, D. O. Jones, A. Rest, {\it et al.}, The Complete Light-curve Sample of Spectroscopically Confirmed SNe Ia from Pan-STARRS1 and Cosmological Constraints from the Combined Pantheon Sample, \href{https://ui.adsabs.harvard.edu/abs/2018ApJ...859..101S}{{\em ApJ} {\bf 859}, 101} (2018)

\bibitem{weinberg}
S. Weinberg, {\em Cosmology} (Oxford University Press, 2008)

\bibitem{Chevallier}
M. Chevallier, and D. Polarski, Accelerating Universes with Scaling Dark Matter, \href{https://ui.adsabs.harvard.edu/abs/2001IJMPD..10..213C}{{\em IJMPD} {\bf 10}, 213} (2001) 

\bibitem{Linder}
E. V. Linder, Exploring the Expansion History of the Universe, \href{https://ui.adsabs.harvard.edu/abs/2003PhRvL..90i1301L}{{\em Phys. Rev. Lett.} {\bf 90}, 091301} (2003)

\bibitem{hu-sawicki}
W. Hu, and I. Sawicki, Models of $f(R)$ cosmic acceleration that evade solar system tests, \href{https://journals.aps.org/prd/abstract/10.1103/PhysRevD.76.064004}{{\em Phys. Rev. D} {\bf 76}, 064004} (2007)

\bibitem{starob}
A. A. Starobinsky, Disappearing cosmological constant in $f(R)$ gravity, \href{https://link.springer.com/article/10.1134\%2FS0021364007150027}{{\em JETP Lett.} {\bf 86}, 157} (2007)

\bibitem{tsujik}
S. Tsujikawa, Observational signatures of $f(R)$ dark energy models that satisfy cosmological and local gravity constraints, \href{https://journals.aps.org/prd/abstract/10.1103/PhysRevD.77.023507}{{\em Phys. Rev. D} {\bf 77}, 023507} (2008)

\bibitem{oesch}
P. A. Oesch, G. Brammer, P. G. van Dokkum, {\it et al.}, A Remarkably Luminous Galaxy at z=11.1 Measured with Hubble Space Telescope Grism Spectroscopy, \href{https://ui.adsabs.harvard.edu/abs/2016ApJ...819..129O}{{\em ApJ} {\bf 819}, 129} (2016)

\bibitem{Kaz}
L. Kazantzidis, and L. Perivolaropoulos, Hints of a local matter underdensity or modified gravity in the low $z$ Pantheon data, \href{https://link.aps.org/doi/10.1103/PhysRevD.102.023520}{{\em Phys. Rev. D} {\bf 102}, 023520} (2020) 

\bibitem{14Cardone2009}
V. F. Cardone, S. Capozziello, and M. G. Dainotti, An updated gamma-ray bursts Hubble diagram, \href{https://academic.oup.com/mnras/article/400/2/775/1016987}{{\em MNRAS} {\bf 400}, 775} (2009)

\bibitem{15Cardone2010}
V. F. Cardone, M. G. Dainotti, S. Capozziello, and R. Willingale, Constraining cosmological parameters by gamma-ray burst X-ray afterglow light curves, \href{https://academic.oup.com/mnras/article/408/2/1181/1028753}{{\em MNRAS} {\bf 408}, 1181} (2010)

\bibitem{17Dainotti2013}
M. G. Dainotti, V. F. Cardone, E. Piedipalumbo, and S. Capozziello, Slope evolution of GRB correlations and cosmology, \href{https://academic.oup.com/mnras/article/436/1/82/970802}{{\em MNRAS} {\bf 436}, 82} (2013)

\bibitem{18Postnikov2014}
S. Postnikov, M. G. Dainotti, X. Hernandez, and S. Capozziello, Nonparametric Study of the Evolution of the Cosmological Equation of State with SNe Ia, BAO, and High-Redshift GRBs, \href{https://iopscience.iop.org/article/10.1088/0004-637X/783/2/126}{{\em ApJ} {\bf 783}, 126} (2014)

\bibitem{361Dainotti2021}
M. G. Dainotti, M. Bogdan, A. Narendra, {\it et al.}, Predicting the Redshift of $\gamma$-Ray-loud AGNs Using Supervised Machine Learning, \href{https://iopscience.iop.org/article/10.3847/1538-4357/ac1748}{{\em ApJ} {\bf 920}, 118} (2021)

\bibitem{362Dainotti2019}
M. G. Dainotti, V. Petrosian, M.  Bogdan, {\it et al.},
Gamma-ray Bursts as distance indicators through a machine learning approach, \href{https://arxiv.org/abs/1907.05074}{{\em arXiv} {\bf 2019}, arXiv:1907.05074}

\bibitem{24Dainotti2008}
M. G. Dainotti, V. F. Cardone, and S. Capozziello, A time-luminosity correlation for $\gamma$-ray bursts in the X-rays, \href{https://academic.oup.com/mnrasl/article/391/1/L79/1127619}{{\em MNRASL} {\bf 391}, L79} (2008)

\bibitem{25Dainotti2010}
M. G. Dainotti, R. Willingale, S. Capozziello, V. F. Cardone, and M. Ostrowski, Discovery of a Tight Correlation for Gamma-Ray Burst Afterglows with “Canonical” Light Curves, \href{https://iopscience.iop.org/article/10.1088/2041-8205/722/2/L215}{{\em ApJL} {\bf 722}, L215} (2010)

\bibitem{26Dainotti2011}
M. G. Dainotti, V. F. Cardone, S. Capozziello, M. Ostrowski, and R. Willingale, Study of possible systematics in the $L^{*}_{X}$ - $T^{*}_a$ correlation of Gamma Ray Bursts, \href{https://iopscience.iop.org/article/10.1088/0004-637X/730/2/135}{{\em ApJ} {\bf 730}, 135} (2011)

\bibitem{27Dainotti2013}
M. G. Dainotti, V. Petrosian, J. Singal, and M. Ostrowski, Determination of the intrinsic luminosity time correlation in the X-ray afterglows of gamma-ray bursts, \href{https://iopscience.iop.org/article/10.1088/0004-637X/774/2/157}{{\em ApJ} {\bf 774}, 157} (2013)

\bibitem{28Dainotti2015}
M. G. Dainotti, R. Del Vecchio, N. Shigehiro, and S. Capozziello, Selection effects in Gamma-Ray Burst Correlations: Consequences on the Ratio Between Gamma-Ray Burst And Star Formation rates, \href{https://iopscience.iop.org/article/10.1088/0004-637X/800/1/31}{{\em ApJ} {\bf 800}, 31} (2015)

\bibitem{29Dainotti2015}
M. G. Dainotti, V. Petrosian, R. Willingale, {\it et al.}, Luminosity–time and luminosity–luminosity correlations for GRB prompt and afterglow plateau emissions, \href{https://academic.oup.com/mnras/article/451/4/3898/1116256}{{\em MNRAS} {\bf 451}, 3898} (2015)

\bibitem{30Dainotti2016}
M. G. Dainotti, S. Postnikov, X. Hernandez, and M. Ostrowski, A fundamental Plane for Long Gamma-Ray Bursts with X-Ray plateaus, \href{https://iopscience.iop.org/article/10.3847/2041-8205/825/2/L20}{{\em ApJL} {\bf 825}, L20} (2016)

\bibitem{31Dainotti2017}
M. G. Dainotti, S. Nagataki, K. Maeda, S. Postnikov, and E. Pian, A study of gamma ray bursts with afterglow plateau phases associated with supernovae, \href{https://www.aanda.org/articles/aa/full_html/2017/04/aa28384-16/aa28384-16.html}{{\em A\&A} {\bf 600}, A98} (2017)

\bibitem{32Dainotti2017}
M. G. Dainotti, X. Hernandez, S. Postnikov, {\it et al.}, A Study of the Gamma-Ray Burst Fundamental Plane, \href{https://iopscience.iop.org/article/10.3847/1538-4357/aa8a6b}{{\em ApJ} {\bf 848}, 88} (2017)

\bibitem{33Dainotti2018}
M. G. Dainotti, and L. Amati,Gamma-ray Burst Prompt Correlations: Selection and Instrumental Effects, \href{https://iopscience.iop.org/article/10.1088/1538-3873/aaa8d7}{{\em PASP} {\bf 130}, 051001} (2018) 

\bibitem{34Dainotti2018}
M. G. Dainotti, R. Del Vecchio, R., and M. Tarnopolski, Gamma-Ray Burst Prompt Correlations, \href{https://www.hindawi.com/journals/aa/2018/4969503/}{{\em Adv. Astron.} {\bf 2018}, 4969503} (2018)

\bibitem{35Dainotti2020}
M. G. Dainotti, A. L. Lenart, G. Sarracino, {\it et al.}, The X-Ray Fundamental Plane of the Platinum Sample, the Kilonovae, and the SNe Ib/c Associated with GRBs, \href{https://iopscience.iop.org/article/10.3847/1538-4357/abbe8a}{{\em ApJ} {\bf 904}, 97} (2020)

\bibitem{36Dainotti2021}
M. G. Dainotti, N. Omodei, G. P. Srinivasaragavan, {\it et al.}, On the Existence of the Plateau Emission in High-energy Gamma-Ray Burst Light Curves Observed by Fermi-LAT, \href{https://iopscience.iop.org/article/10.3847/1538-4365/abfe17}{{\em ApJS} {\bf 255}, 13} (2021)

\bibitem{37Dainotti2021}
M. G. Dainotti, D. Levine, N. Fraija, and P. Chandra, Accounting for Selection Bias and Redshift Evolution in GRB Radio Afterglow Data, \href{https://www.mdpi.com/2075-4434/9/4/95}{{\em Galaxies} {\bf 9}, 95} (2021)

\bibitem{38DelVecchio2016}
R. Del Vecchio, M. G. Dainotti, and M. Ostrowski, Study of GRB Light-Curve Decay Indices in the Afterglow phase, \href{https://iopscience.iop.org/article/10.3847/0004-637X/828/1/36}{{\em ApJ} {\bf 828}, 36} (2016) 

\bibitem{349Dainotti2021}
M. G. Dainotti, V. Petrosian, and L. Bowden, Cosmological Evolution of the Formation Rate of Short Gamma-Ray Bursts with and without Extended Emission, \href{https://iopscience.iop.org/article/10.3847/2041-8213/abf5e4}{{\em ApJ} {\bf 914}, L40} (2021)

\bibitem{353Cao2022}
S. Cao, M. G. Dainotti, and B. Ratra, Standardizing Platinum Dainotti-correlated gamma-ray bursts, and using them with standardized Amati-correlated gamma-ray bursts to constrain cosmological model parameters, \href{https://arxiv.org/abs/2201.05245}{{\em arXiv} {\bf 2022}, arXiv:astro-ph.CO/2201.05245}

\bibitem{41Rowlinson2014}
A. Rowlinson, B. P. Gompertz, M. G. Dainotti, {\it et al.}, Constraining properties of GRB magnetar central engines using the observed plateau luminosity and duration correlation, \href{https://academic.oup.com/mnras/article/443/2/1779/1070941}{{\em MNRAS} {\bf 443}, 1779} (2014)

\bibitem{42Rea2015}
N. Rea, M. Gullón, J. A. Pons, {\it et al.}, Constraining The GRB-Magnetar Model by Means of the Galactic Pulsar Population, \href{https://iopscience.iop.org/article/10.1088/0004-637X/813/2/92}{{\em ApJ} {\bf 813}, 92} (2015)

\bibitem{43Stratta2018}
G. Stratta, M. G. Dainotti, S. Dall’Osso, X. Hernandez, and G. De Cesare, On the Magnetar Origin of the GRBs Presenting X-Ray Afterglow Plateaus, \href{https://iopscience.iop.org/article/10.3847/1538-4357/aadd8f}{{\em ApJ} {\bf 869}, 155} (2018)





\end{thebibliography}

\end{document}